\documentclass[aps,prb,reprint,twocolumn,superscriptaddress,floatfix,nofootinbib]{revtex4-1}
\usepackage{amsmath,amssymb,color,comment,physics}
\usepackage[makeroom]{cancel}
\usepackage{mathrsfs}
\usepackage{graphicx}
\usepackage[countmax]{subfloat}
\usepackage[english]{babel}
\definecolor{jadclr}{rgb}{0,0.5,0}
\definecolor{jadcolor}{rgb}{0.5,0,0}
\usepackage[bookmarks=true,colorlinks,linkcolor=jadclr,urlcolor=jadcolor,citecolor=jadcolor]{hyperref}
\usepackage{subfigure}
\usepackage{dsfont}

\def\d{\mathrm d}

\definecolor{mypink1}{rgb}{0.858, 0.188, 0.478}

\begin{document}
\title{Dynamical criticality and domain-wall coupling in long-range Hamiltonians}

\author{Nicol\`o Defenu}
\affiliation{Institut f\"ur Theoretische Physik, Universit\"at Heidelberg, D-69120 Heidelberg, Germany}

\author{Tilman Enss}
\affiliation{Institut f\"ur Theoretische Physik, Universit\"at Heidelberg, D-69120 Heidelberg, Germany}


\author{Jad C.~Halimeh}
\affiliation{Max Planck Institute for the Physics of Complex Systems, N\"othnitzer Stra\ss e 38, 01187 Dresden, Germany}
\affiliation{Physics Department, Technical University of Munich, 85747 Garching, Germany}

\begin{abstract}
Dynamical quantum phase transitions hold a deep connection to the underlying equilibrium physics of the quench Hamiltonian. In a recent study [J.~C.~Halimeh \textit{et al.}, arXiv:1810.07187], it has been numerically demonstrated that the appearance of anomalous cusps in the Loschmidt return rate coincides with the presence of bound domain walls in the spectrum of the quench Hamiltonian. Here, we consider transverse-field Ising chains with power-law and exponentially decaying interactions, and show that by removing domain-wall coupling via a truncated Jordan-Wigner transformation onto a Kitaev chain with long-range hopping and pairing, anomalous dynamical criticality is no longer present. This indicates that bound domain walls are necessary for anomalous cusps to appear in the Loschmidt return rate. We also calculate the dynamical phase diagram of the Kitaev chain with long-range hopping and pairing, which in the case of power-law couplings is shown to exhibit rich dynamical criticality including a doubly critical dynamical phase.
\end{abstract}

\date{\today}
\maketitle
\section{Introduction}
After the establishment of the theoretical framework for equilibrium thermal and quantum phase transitions\cite{Cardy_book,Ma_book,Sachdev_book} including the vastly successful renormalization group method,\cite{Wilson1971a,Wilson1971b,Wilson1972,Wilson1974,Wilson1975} out-of-equilibrium criticality has become an ever growing branch of physics. In classical systems, the concepts of dynamical universality, scaling, and renormalization group have been developed.\cite{Hohenberg1977} In quantum many-body systems, such concepts are still not fully understood. In recent years, two seemingly disparate concepts of dynamical phase transitions have received a lot of attention, both of which entail a \textit{quantum quench} protocol where a control parameter is instantaneously changed in the system. The first concept relies on a local order parameter\cite{footnote1} in much the same way as in the Landau theory of equilibrium phase transitions.\cite{Cardy_book} After a quantum quench, the value of the order parameter of the long-time steady state determines its dynamical phase. This has been studied in mean-field models,\cite{Sciolla2010,Sciolla2011,Sciolla2013,Homrighausen2017,Lang2017,Lang2018} nonintegrable transverse-field Ising chains,\cite{Halimeh2017b,Halimeh2017,Zunkovic2018} and the two-dimensional quantum Ising model.\cite{Hashizume2018}

The second concept is the so-called \textit{dynamical quantum phase transitions} (DQPT),\cite{footnote1} which have been established in recent years as an intuitive way to classify criticality out of equilibrium by defining a dynamical analog of the thermal free energy.\cite{Heyl2013} In statistical mechanics, nonanalyticities in the thermal free energy indicate a thermal phase transition whose order equals that of the free-energy temperature derivative in which there is a discontinuity, as per the Ehrenfest classification.\cite{Ma_book,Cardy_book} In out-of-equilibrium quantum many-body physics, the dynamical analog of the thermal partition function is the Loschmidt amplitude, which is the overlap of the time-evolved state with its initial self in the wake of a quantum quench. The Loschmidt amplitude is itself a boundary partition function, where complexified time can be construed as inverse temperature.\cite{Heyl2013,Heyl_review} The Loschmidt return rate, defined as the negative of the logarithm of the Loschmidt amplitude, is then the dynamical free energy density, and its nonanalyticities occur at specific \textit{critical times}. In the seminal work of Ref.~\onlinecite{Heyl2013}, it is shown in the nearest-neighbor transverse-field Ising chain (NNTFIC) that only quenches across the equilibrium critical point lead to nonanalyticities in the form of cusps occurring at equally spaced times in every cycle of the return rate, where these cusps are connected to a single critical momentum mode in the Jordan-Wigner (JW) fermionic basis. Additionally, if the quench starts off in the ordered phase, these \textit{regular} cusps hold a direct connection to and the same periodicity as zero crossings in the order parameter.

The picture drastically changes in the presence of extensive long-range interactions\cite{footnote3} such as power-law\cite{Halimeh2017,Zauner2017,Zunkovic2018} and exponentially decaying\cite{Halimeh2018a} profiles in one-dimensional (1D) quantum Ising chains, where at small enough final transverse-field strength a distinct kind of \textit{anomalous} cusps arise in the return rate that are neither connected to zero crossings of the order parameter nor are they necessarily spaced at equal time intervals.\cite{Halimeh2017} Besides their characteristic property of appearing in the return rate even when the order parameter makes no zero crossings, anomalous cusps are distinguished from their regular counterparts in that they have been shown to belong to a separate group of Fisher zeros,\cite{Zauner2017} and they are related to the existence of bound domain walls in the spectrum of the quench Hamiltonian.\cite{Halimeh2018a} In two-dimensional systems, anomalous cusps occur even in the case of nearest-neighbor interactions.\cite{Hashizume2018} In the fully connected transverse-field Ising model (FCTFIM), anomalous cusps arise for sufficiently small quenches starting in any ordered thermal state.\cite{Homrighausen2017,Lang2017,Lang2018}

Recently, the connection of DQPT and the quasiparticle spectrum of the quench Hamiltonian have received a lot of attention.\cite{Halimeh2018a,Hashizume2018,Jafari2019} In Ref.~\onlinecite{Halimeh2018a}, numerical evidence strongly supports a connection between bound domain walls in the spectrum of the quench Hamiltonian and anomalous cusps in 1D quantum Ising chains with extensive long-range interactions. Specifically, when local spin excitations are energetically favorable to two-domain-wall states in the spectrum of the quench Hamiltonian, anomalous cusps will appear in the return rate. This additionally coincides with a long-lived prethermal state where the order parameter exhibits very slow decay even when the model has no finite-temperature phase transition.\cite{Liu2018,Halimeh2018a} Domain walls do not form bound states in NNTFIC, but in certain quantum spin chains with extensive interactions, domain-wall binding can occur for sufficiently small transverse-field strength, with the crossover value of the latter growing larger the more extended the interaction range is. Whether or not bound domain walls are a necessary condition for the existence of anomalous cusps has not yet been addressed. In this work, we show that after removing domain-wall coupling in the long-range transverse-field Ising chain (LRTFIC) by mapping it using a \textit{truncated JW transformation} onto the long-range Kitaev chain (LRKC), anomalous cusps no longer appear in the return rate. Nevertheless, DQPT in LRKC with sufficiently long-range power-law hopping and pairing exhibit critical behavior distinct from that of NNTFIC in that the former supports a \textit{doubly critical} regular dynamical phase in the wake of certain quenches where cusps appear in the return rate due to \textit{two} critical momentum modes, such that each set of cusps exhibits its own periodicity. For exponentially decaying pairing and hopping, one finds the two traditional dynamical phases of either cusps due to a single critical momentum mode (\textit{singly critical} regular phase) or no cusps at all (trivial phase).

The remainder of our paper is structured as follows: In Sec.~\ref{sec:model}, we introduce LRTFIC and perform the truncated JW mapping onto LRKC. In Sec.~\ref{sec:dynamics}, we analytically derive the quench dynamics for the return rate, and construct the dynamical phase diagram for LRKC in the cases of power-law and exponentially decaying hopping and pairing. We interpret our results and how they contrast those of LRTFIC from the perspective of domain walls and their coupling in Sec.~\ref{sec:dw}. We conclude in Sec.~\ref{sec:conc} and provide further details of our derivation in Appendix~\ref{AppB}.

\section{Model and mapping to fermions}\label{sec:model}
We shall consider LRTFIC described by the Hamiltonian
\begin{align}\label{eq:LRTFIC}
\hat{H}=-\sum_{l<j}V_{|l-j|}\hat{\sigma}^z_l\hat{\sigma}^z_j-h\sum_j\hat{\sigma}^x_j,
\end{align}
where $\hat{\sigma}^{\{x,y,z\}}_j$ are the Pauli spin matrices on site $j$, $h$ is the transverse-field strength, and $V_{r}$ is the spin coupling profile that in this work shall be either power-law ($\propto1/r^\alpha$, $\alpha\geq0$) or exponential  ($\propto\exp[-\lambda(r-1)]$, $\lambda\geq0$) decay. In the limit $\alpha,\lambda\to\infty$, Hamiltonian~\eqref{eq:LRTFIC} represents NNTFIC, which is integrable and can be exactly solved with a JW transformation.\cite{Sachdev_book} In the other integrable limit $\alpha,\lambda=0$,~\eqref{eq:LRTFIC} describes FCTFIM, which is also known as the Lipkin-Meshkov-Glick (LMG) model.\cite{Lipkin1965,Meshkov1965,Glick1965} In FCTFIM, permutation symmetry due to infinite-range interactions allows mapping the problem onto a Dicke basis that is linear in system size (rather than exponential as in the original basis), which in turn makes the system amenable to exact diagonalization. For power-law interactions with $\alpha<2$, Hamiltonian~\eqref{eq:LRTFIC} supports a finite-temperature phase transition,\cite{Dyson1969,Thouless1969,Dutta2001} which belongs to the same universality class of the classical long-range Ising model.\cite{Defenu2015} At zero temperature, LRTFIC hosts a second-order quantum phase transition. At mean-field level, depending on the value of $\alpha,\lambda$, the equilibrium universality changes from that of traditional nearest-neighbor interactions ($\alpha\geq3,\lambda>0$) to the long-range correlated regime for $\alpha\in(5/3,3)$, and it finally reaches mean-field behavior for $\alpha<5/3,\lambda=0$.\cite{Maghrebi2016, Defenu2017} When fluctuations beyond mean-field are introduced, the above picture remains unchanged apart from the boundary between long-range and nearest-neighbor universalities, which is shifted from $\alpha=3$ to a value $\alpha^{*}<3$.\cite{Defenu2017} It is worth noting that the universal behavior of long-range systems can be, at least approximately, connected to the one of their nearest-neighbor equivalent in real fractional dimensions.\cite{Defenu2015,Defenu2016,Defenu2017, Gori2017} DQPT in Hamiltonian~\eqref{eq:LRTFIC} have been studied in both the integrable \cite{Heyl2013,Homrighausen2017,Lang2017,Lang2018} and nonintegrable \cite{Halimeh2017,Zauner2017} cases. Anomalous dynamical criticality has been well established in the case of sufficiently long-range interactions, \cite{Homrighausen2017,Halimeh2017,Zauner2017,Lang2017,Lang2018,Halimeh2018a,Halimeh2019} and has been connected to the presence of bound domain walls in the spectrum of the quench Hamiltonian.\cite{Halimeh2019} Moreover, the universal slow dynamics of LRTFIC has been extensively studied both in the nearest-neighbor ($\alpha,\lambda\to\infty$) and fully connected ($\alpha,\lambda=0$) limits.\cite{Dziarmaga2005,Karl2017,Defenu2018}

Qualitative understanding of LRTFIC can be achieved by mapping~\eqref{eq:LRTFIC} onto fermions using the JW transformation\cite{Jaschke2017,Vanderstraeten2018}
\begin{align}
\hat{\sigma}_j^x&=1-2\hat{c}_j^\dagger\hat{c}_j,\\
\hat{\sigma}_j^y&=-\mathrm{i}\Big[\prod_{m=1}^{j-1}\big(1-2\hat{c}_m^\dagger\hat{c}_m\big)\Big]\big(\hat{c}_j-\hat{c}_j^\dagger\big),\\
\hat{\sigma}_j^z&=-\Big[\prod_{m=1}^{j-1}\big(1-2\hat{c}_m^\dagger\hat{c}_m\big)\Big]\big(\hat{c}_j+\hat{c}_j^\dagger\big),
\end{align}
where $\hat{c}_j,\hat{c}_j^\dagger$ are fermionic annihilation and creation operators, respectively, that satisfy the canonical anticommutation relations $\{\hat{c}_l,\hat{c}_j\}=0$ and $\{\hat{c}_l,\hat{c}_j^\dagger\}=\delta_{l,j}$. This renders~\eqref{eq:LRTFIC} in the fermionic form
\begin{align}\nonumber
\hat{H}=&\,-\sum_{l<j}V_{|l-j|}\big(\hat{c}_l^\dagger-\hat{c}_l\big)\Big[\prod_{n=l+1}^{j-1}\big(1-2\hat{c}_n^\dagger\hat{c}_n\big)\Big]\big(\hat{c}_j^\dagger+\hat{c}_j\big)\\\label{eq:quartic}
&-h\sum_j\big(1-2\hat{c}_j^\dagger\hat{c}_j\big).
\end{align}
The Hamiltonian~\eqref{eq:quartic} cannot be exactly solved, due to the presence of higher-than-quadratic-order terms in the fermionic operators. We employ the approximation
\begin{align}\label{eq:approx}
\prod_{n=l+1}^{j-1}\big(1-2\hat{c}_n^\dagger\hat{c}_n\big)=1,
\end{align}
for every $j\geq l+2$, neglecting the string operators in the interaction terms in the first line of~\eqref{eq:quartic}. This \textit{truncated} JW transformation leads to the quadratic Hamiltonian
\begin{align}\nonumber
\hat{H}=&\,-\sum_{l<j}V_{|l-j|}\big(\hat{c}_l^\dagger\hat{c}_j+\hat{c}_l^\dagger\hat{c}_j^\dagger-\hat{c}_l\hat{c}_j-\hat{c}_l\hat{c}_j^\dagger\big)\\
&-h\sum_j\big(1-2\hat{c}_j^\dagger\hat{c}_j\big),
\label{lrk_h}
\end{align}
which we shall refer to as the long-range Kitaev chain (LRKC), as in the limit $\alpha,\lambda\to\infty$ it is the paradigmatic Kitaev chain at equal nearest-neighbor hopping and pairing strengths. Kitaev chains in presence of long-range pairing and hopping terms have already been studied in the equilibrium context,\cite{Viyuela2015,Viyuela:2018fpv} showing that for slow enough decay rates long-range pairing effects alter the nature of the topological phase\cite{Vodola2014,Lepori2015,Lepori2017And, Alecce2017} and the spreading of correlations.\cite{Foss-Feig2015,Vodola2016}
The approximation~\eqref{eq:approx} is equivalent to a domain-wall Hamiltonian in the spin basis, where interaction terms between domain walls caused by long-range interactions have been removed. Indeed, the single fermionic states introduced by the JW transformation are equivalent to domain-wall configurations when represented in the spin basis;\cite{Fradkin2017} cf.~Sec.~\ref{sec:dw} for mathematical details.

The Hamiltonian~\eqref{lrk_h} is translation invariant and is thus more conveniently represented in Fourier space as
\begin{align}
\label{h_klr}
\hat{H}=\int\d k\left[(\hat{c}^{\dagger}_{k}\hat{c}_{k}
-\hat{c}_{-k}\hat{c}^{\dagger}_{-k})\varepsilon_{k}+(\hat{c}^{\dagger}_{k}\hat{c}^{\dagger}_{-k}+\hat{c}_{-k}\hat{c}_{k})\Delta_{k}\right],
\end{align}
where the exact expressions for $\varepsilon_k$ and $\Delta_k$ are given in Appendix~\ref{AppB} for both power-law and exponentially decaying hopping and pairing. Hamiltonian~\eqref{h_klr} can be diagonalized by a Bogoliubov transformation, the details of which are also presented in Appendix~\ref{AppB}. The ground state of the system is the BCS ground state
\begin{align}
|\Psi_{0}\rangle=\prod_{k}\left(\cos\frac{\theta_{k}}{2}+\sin\frac{\theta_{k}}{2}\hat{c}^{\dagger}_{k}\hat{c}^{\dagger}_{-k}\right)|0\rangle,
\end{align}
where $\tan\theta_{k}=\Delta_{k}/\varepsilon_{k}$ and $\ket{0}$ is the vacuum state.

\begin{figure*}[t!]
	\centering
	\subfigure[$\quad\alpha=1.5$]{\label{Fig1a}\includegraphics[width=.32\textwidth]{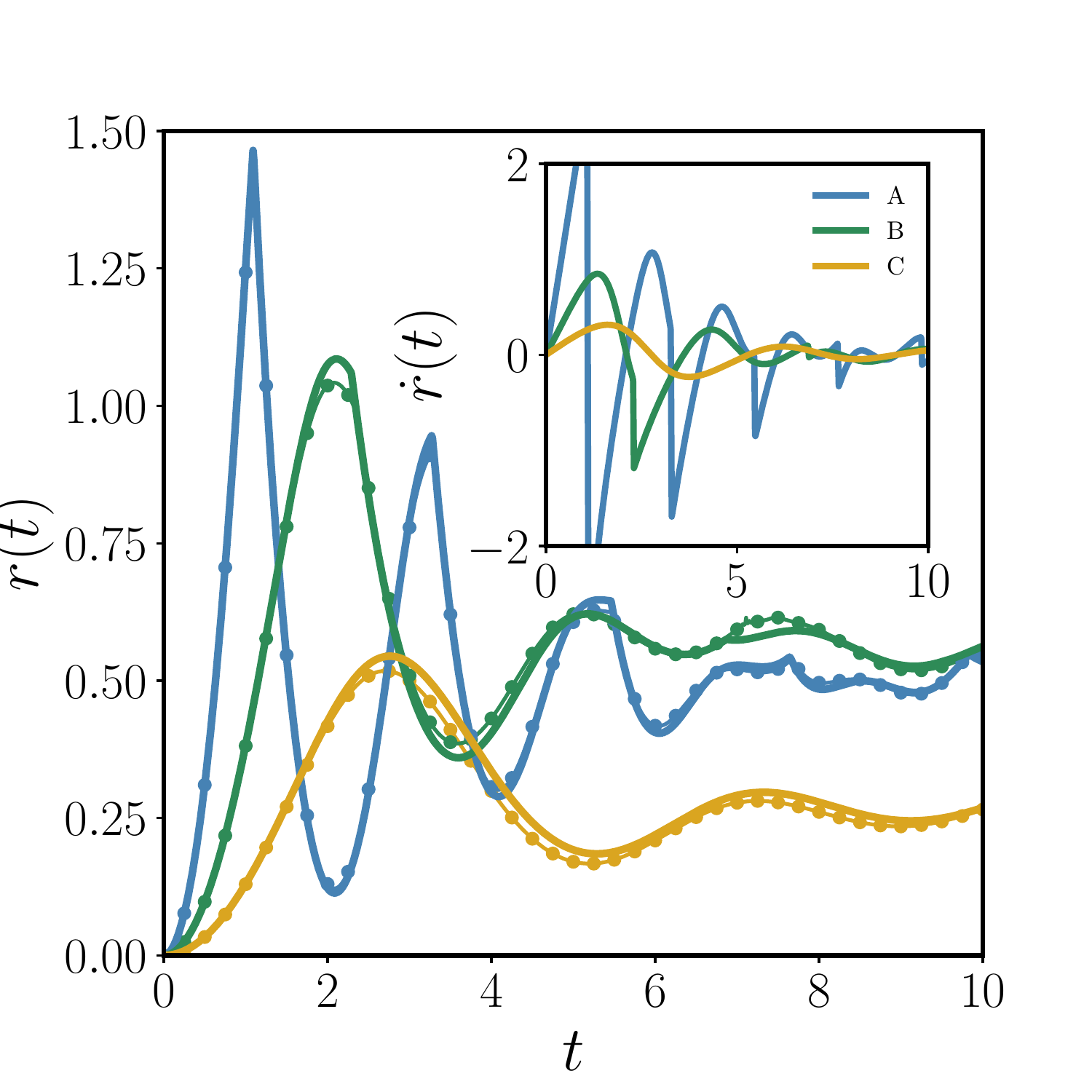}}
	\hfill
	\subfigure[$\quad\alpha=2.1$]{\label{Fig1b}\includegraphics[width=.32\textwidth]{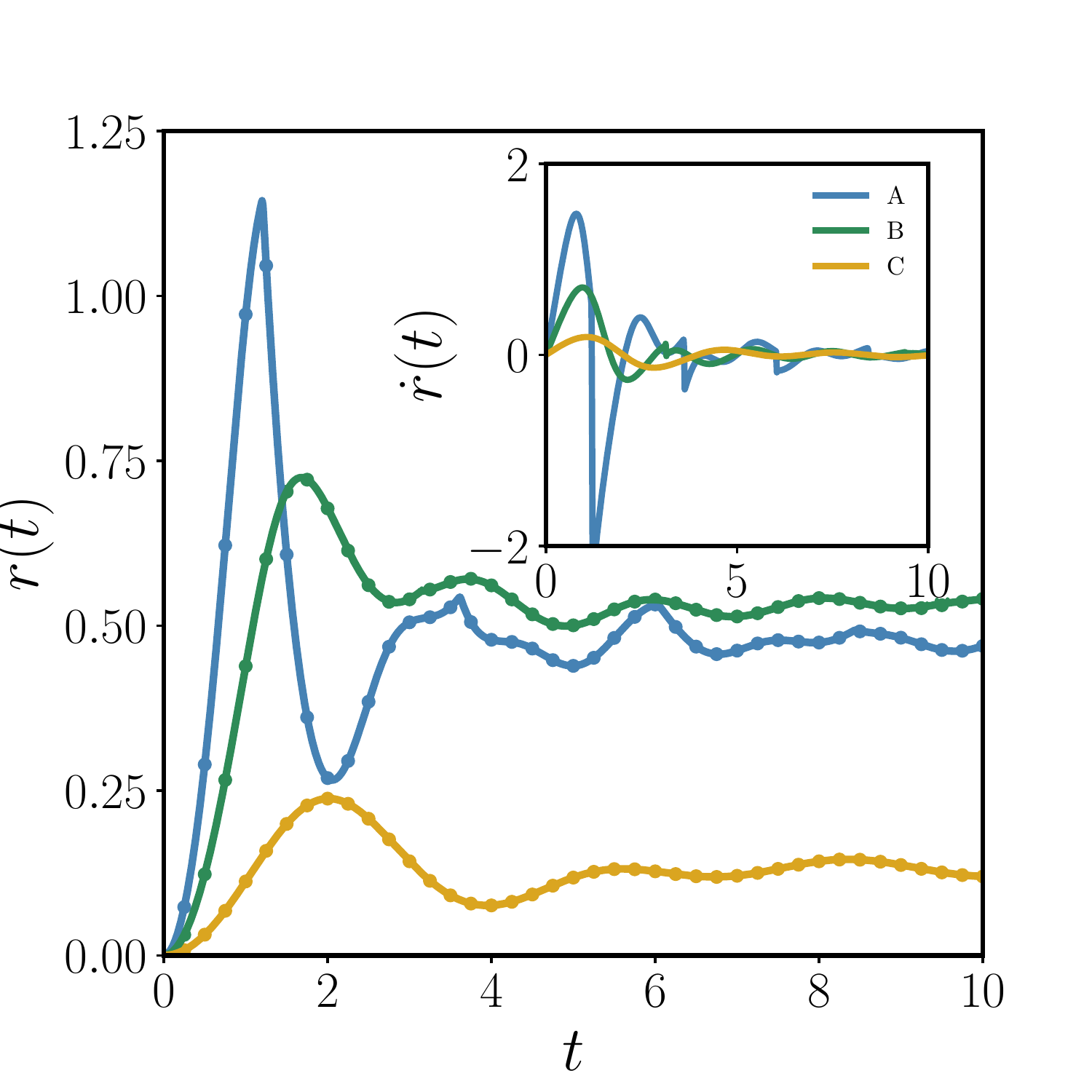}}
	\hfill
	\subfigure[$\quad\alpha=5.5$]{\label{Fig1c}\includegraphics[width=.32\textwidth]{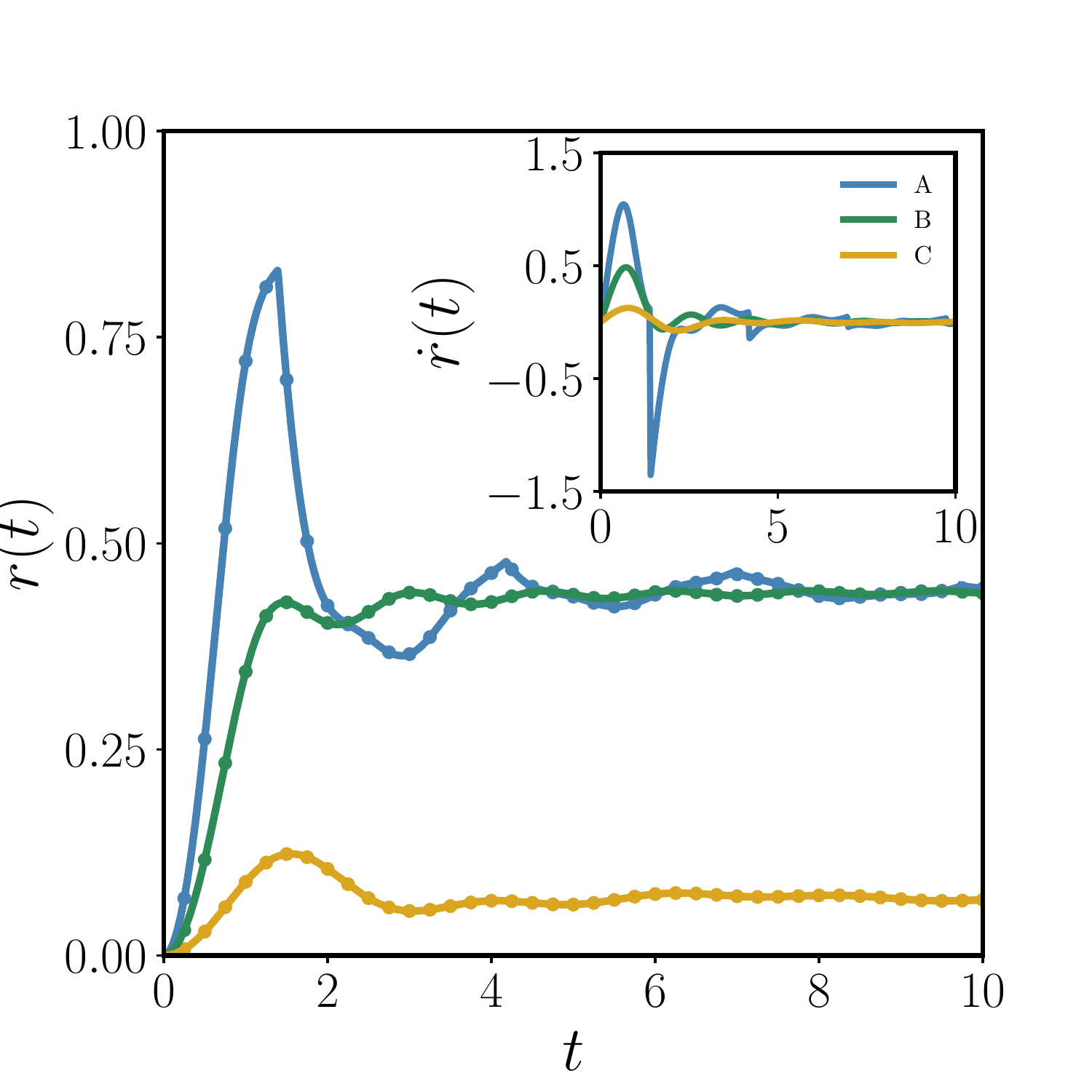}}
	\caption{\label{Fig1} The return rate for LRKC at three values of $\alpha=1.5$, $2.1$, and $5.5$ in panels (a), (b), and (c), respectively, is shown after a quench starting at $h_{i}=0$. Each curve represents a different final magnetic field $h_{f}$ located in the regions A, B, and C (see Fig.~\ref{Fig2}), respectively, from top to bottom. Quenches to region A display the traditional dynamical critical behavior, with singly critical regular cusps appearing when the system instantaneously crosses the equilibrium critical point $h_{c}^{e}=1$; $h_{f}=1.5$ for the upper (blue) curves. The intermediate (green) curves represent the return rate for quenches in region B, corresponding to $h_{f}=0.85$, $0.975$, and $1.0$ for panels (a), (b), and (c), respectively. In panels (a) and (b), these curves show two families of regular cusps originating from two dynamical critical momenta displayed in Fig.~\ref{Fig2b}, while for $\alpha>3$ in panel (c), the region B is collapsed onto a critical line at $h_c^d=h_c^e$, and only a single critical momentum mode exists. Finally, the lower (gold) curves do not show any cusps since the system is quenched to the trivial phase, $h_{f}=0.5$, i.e., to the deep ordered phase below $h_c^d$ at the given $\alpha$ values. In the case of LRTFIC, anomalous cusps appear in this region.\cite{Halimeh2017} In all panels, solid thick curves represent the exact solution in the thermodynamic limit, while dotted-line curves are for a finite chain of length $N=1000$ with periodic boundary conditions; see Appendix~\ref{AppB}. Insets show the time derivative of the return rate, indicating jumps at the critical times.}
\end{figure*}

\section{Quench Dynamics}\label{sec:dynamics}
The system is prepared in the ground state $\ket{\Psi_i}$ of the initial Hamiltonian $\hat{H}_{i}$, which is~\eqref{lrk_h} with transverse-field strength $h_{i}$. At $t=0$, the Hamiltonian is suddenly quenched to $\hat{H}_{f}$, which is with the transverse-field strength $h_{f}$. The ensuing dynamics is encapsulated in the Loschmidt amplitude
\begin{align}
G(t)=\bra{\Psi_i}\mathrm{e}^{-\mathrm{i}\hat{H}_{f}t}\ket{\Psi_i}.
\end{align}
Each momentum state of the system is initially in the ground state $|g_{k}^{i}\rangle$ of the initial Hamiltonian, and it can be expressed as a linear combination of the two states:
\begin{align}
|g_{k}^{i}\rangle=&\,\cos\frac{\theta_k^f-\theta_k^i}{2}|g^f_{k}\rangle+\sin\frac{\theta_k^f-\theta_k^i}{2}|e^f_{k}\rangle,
\end{align}
where $|g^f_{k}\rangle$ and $|e^f_{k}\rangle$ are the ground and excited states of each momentum mode of the final Hamiltonian $\hat{H}_{f}$, respectively, with $\theta_{k}^{i(f)}$ being the Bogoliubov angle of the initial (final) Hamiltonian; cf.~Appendix~\ref{AppB}. 
To describe the quench dynamics, it is useful to look at the probability,
\begin{align}
p_{k}=\big|\langle e^f_{k} |g_{k}^{i}\rangle\big|^{2}=\sin^{2}\bigg(\frac{\theta_{k}^{f}-\theta_{k}^{i}}{2}\bigg),
\end{align}
for each momentum mode to end up in the excited state of the final Hamiltonian after the sudden quench. The return rate at each
instant in time is given exactly by the expression
\begin{align}
\label{exp_ret_rate}
r(t)=-\int_{0}^{\pi}\frac{dk}{2\pi}\ln\big[1+4p_{k}(p_{k}-1)\sin^{2}(\omega_{k}^{f}t)\big],
\end{align}
where $\omega_{k}^{f}$ is the Bogoliubov spectrum in~\eqref{spec_eq} of the final Hamiltonian. It is clear from~\eqref{exp_ret_rate} that nonanalyticities can only occur at critical momenta $k^*$ where $p_{k^*}=1/2$ and at critical times $t^*=(n+1/2)\pi/\omega_{k^*}^f$.

\subsection{Power-law couplings}
\begin{figure*}[t!]
	\centering
	\subfigure[$\quad h_{f}>h_{c}^e$]{\label{Fig2a}\includegraphics[width=.32\textwidth]{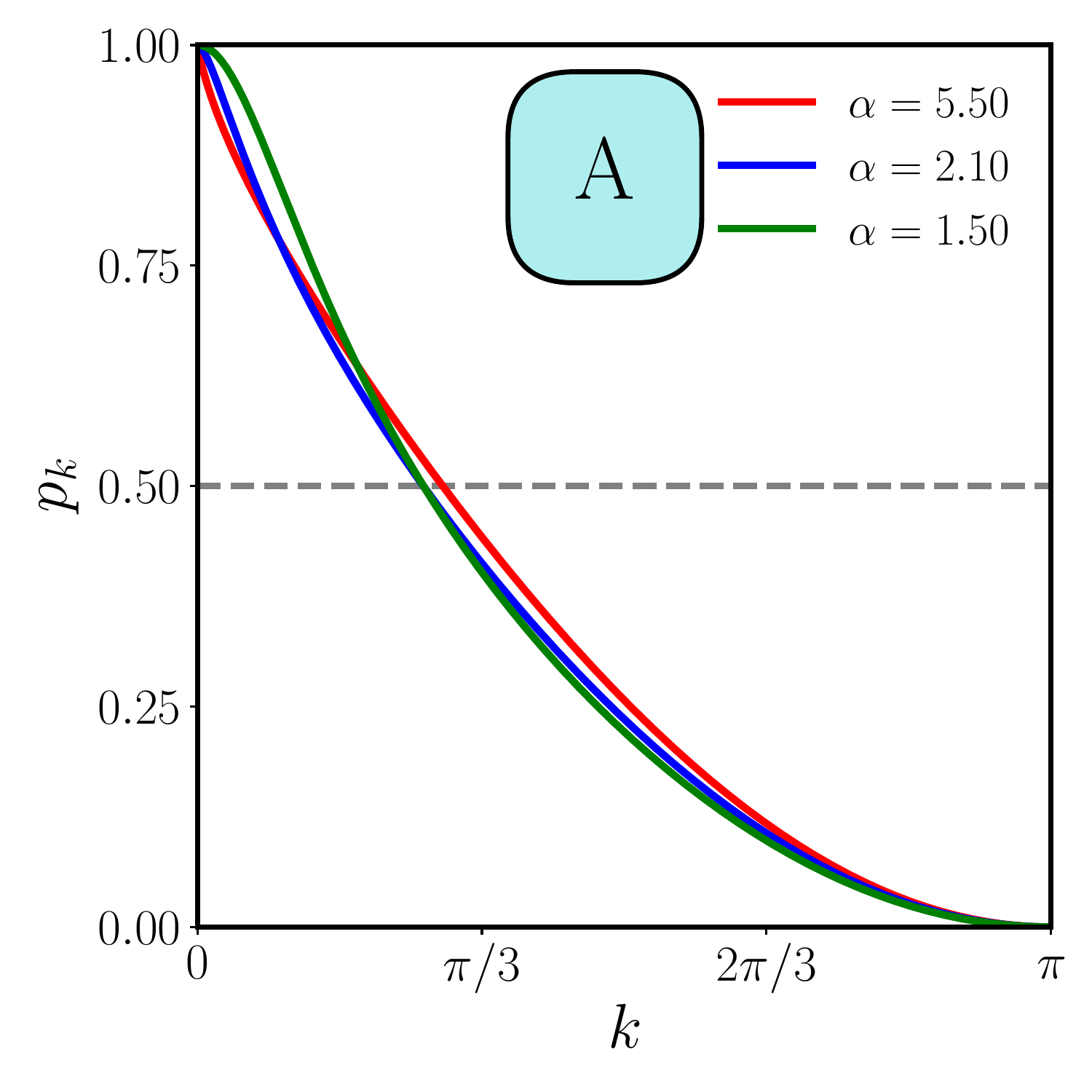}}
	\hfill
	\subfigure[$\quad h_{c}^d<h_f<h_{c}^e$]{\label{Fig2b}\includegraphics[width=.32\textwidth]{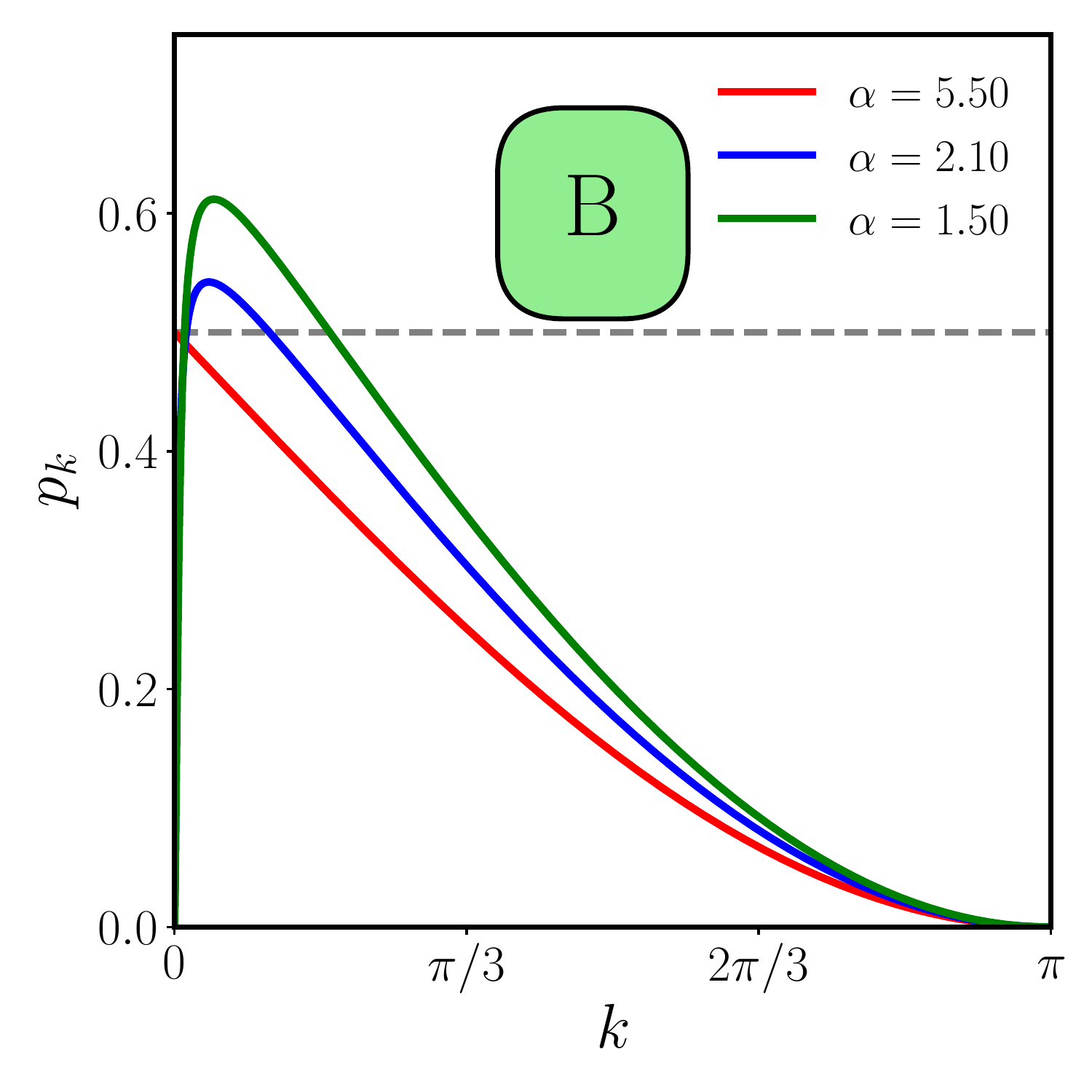}}
	\hfill
	\subfigure[$\quad$Dynamical phase diagram ]{\label{Fig2c}\includegraphics[width=.34\textwidth]{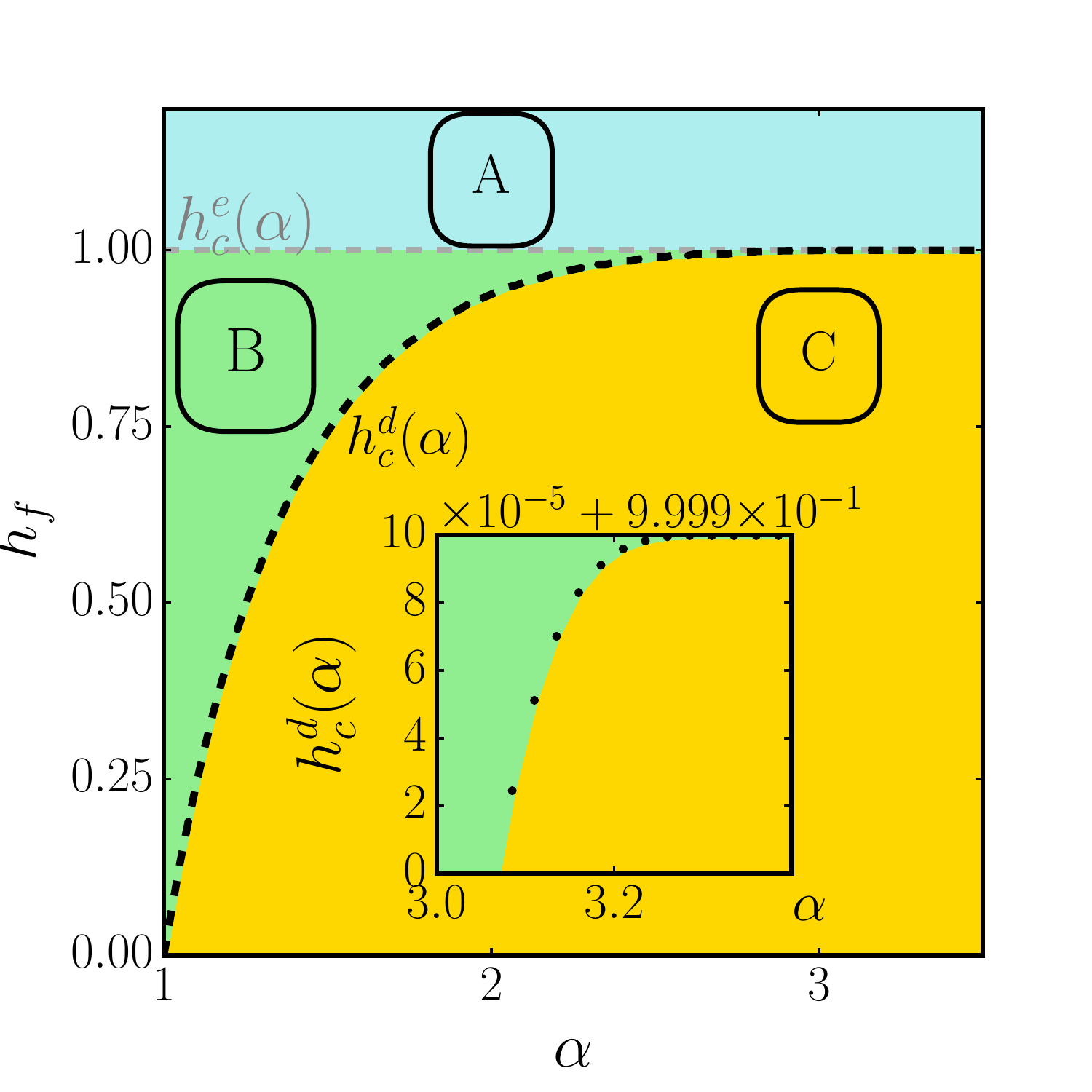}}
	\caption{\label{Fig2} 
		Excitation probabilities for the quenches in Fig.~\ref{Fig1} from $h_i=0$ to final magnetic fields $h_f>h_{c}^e$ (region A) and $h_{c}^d<h_f<h_{c}^e$ (region B), shown in panels (a) and (b), respectively. Panel (c) shows the phase boundaries $h_{c}^{e}(\alpha)$ and $h_{c}^{d}(\alpha)$ as a function of the range parameter $\alpha$. For quenches terminating in region A, one has the singly critical regular dynamical phase known from NNTFIC with a single critical mode $k^{*}$. When $h_{f}$ lies in region B, however, the return rate displays singularities corresponding to two different critical modes, as shown in panel (b), which is only possible for $\alpha<\alpha_{\cross}$ (see text). This doubly critical regular phase vanishes for $\alpha\geq \alpha_{\cross}$ since then $h_c^d=h_c^e$. Finally, the return rate is fully analytic in region C, where no dynamical critical momentum modes exist. In the inset of panel (c), the value of $h_{c}^{d}$ for $\alpha\gtrsim3$ is shown to be very close, but not equal to $h_{c}^{e}=1$.}
\end{figure*}

Let us now consider power-law pairing and hopping, which is equivalent to setting $V_{|l-j|}=|l-j|^{-\alpha}/\mathcal{N}_\alpha$ in~\eqref{lrk_h}, with $\mathcal{N}_\alpha$ the Kac normalization\cite{Kac1963} explicitly given in Appendix~\ref{AppB}. We restrict our discussion to the case of $\alpha>1$. In the nearest-neighbor limit ($\alpha\to\infty$), singly critical regular cusps arise in the return rate for quenches across the equilibrium quantum critical point $h_{c}^e=1$, and no cusps appear for quenches within the same phase.\cite{Heyl2013} For simplicity and without loss of generality, let us restrict our discussion to $h_i=0$. The return rate after various quenches is shown in Fig.~\ref{Fig1} for several values of $\alpha$. At small enough $\alpha$ (to be specified later), a new dynamical critical phase exists for quenches ending in the range $h_{c}^{d}<h_{f}<h_{c}^{e}$, where $h_{c}^{d}$ is an $\alpha$-dependent dynamical critical field. In this phase the return rate shows two different sets of regular singularities, occurring at two different frequencies. We shall call this phase the doubly critical regular dynamical phase. For large enough $\alpha$ (also to be specified later), $h_c^d=h_c^e=1$ and the dynamical criticality of LRKC is that of the nearest-neighbor limit where only one critical mode can exist for quenches to $h_f>h_c^d=h_c^e$. This is the singly critical regular dynamical phase. Indeed, it is possible to classify the two regular dynamical phases by identifying the number of critical modes $k^{*}$ responsible for the nonanalytic behavior of the return rate. According to~\eqref{exp_ret_rate}, critical modes are the ones with equal probability to be either in the ground or excited states, i.e., $p_{k}=1/2$. Looking at the explicit form of the Bogoliubov angle $\theta_{k}=\arctan(\Delta_k/\varepsilon_k)$, it is straightforward to show that the critical modes satisfy the relation
\begin{align}
\label{kstar_eq}
\big[h_{i}-j_{k^{*}}\big]\big[h_{f}-j_{k^{*}}\big]+\Delta_{k^{*}}^{2}=0,
\end{align}
where $j_{k}$ is defined in~\eqref{par_tf_N}. For $h_f>h_{c}^e$ the zero mode is frozen and $p_{k\approx0}\approx1$, while at high energy the dynamics always remains adiabatic $p_{k\approx\pi}\approx0$, and, therefore,~\eqref{kstar_eq} must have at least one solution $0<k^{*}<\pi$, which is responsible for the occurrence of cusps in the return rate; cf.~Fig.~\ref{Fig2a}. On the other hand, for $h_f<h_{c}^e$, we have $p_{k\approx0}\approx0$ since the zero mode is also adiabatic and we expect the excitation probability to remain below $1/2$. However, when long-range interactions are present this is not the case, since there exist a value $h_{c}^d$ above which two solutions arise for~\eqref{kstar_eq}, leading to the appearance of a double crossing excitation probability; see Fig.~\ref{Fig2b}. 
Nevertheless, one can always find a nonzero range of values $h_f\in(0,h_c^d)$ at which the return rate is fully analytic regardless of $\alpha$, because $h_c^d>0$ for $\alpha>1$. This means that, unlike in the case of LRTFIC,\cite{Halimeh2017,Zauner2017} anomalous cusps never arise in the return rate in the case of LRKC. Note how for quenches to $h_{f}>h_{c}^e$ the return rate exhibits sharp cusps at its maxima, as occurs in the upper (blue) curves of Fig.~\ref{Fig1} at $h_{f}=1.5$ for $\alpha=1.5$, $2.1$, and $5.5$. However, for quenches in the region $h_{c}^d<h_{f}<h_{c}^e$ the return rate has \textit{shoulder} singularities following a smooth maximum; see the intermediate (green) curves in Fig.~\ref{Fig1}, panels (a) and (b), where $h_{f}=0.85$ and $0.975$, respectively. These singularities are less evident than those for $h_f>h_{c}^e$, but they can be easily identified by looking at the time derivative $\dot{r}(t)$ of the return rate, shown in the insets of Fig.~\ref{Fig1}; see also the excitation probability $p_{k}$ in Fig.~\ref{Fig2}.

It is worth noting that dynamical phases similar to, albeit not the same as, the doubly critical regular phase occurring in the region $h_{c}^d<h_{f}<h_{c}^e$ have also been found in the dynamics of the Kitaev chain with long-range pairing only (while hopping is nearest-neighbor), where these phases contain three critical momenta, rather than two, and exist only for $\alpha<2$. However, no disparity in range between pairing and hopping terms exists in either the fermionic representation of the Ising model, see~\eqref{eq:quartic}, nor in the truncated JW approximation, see~\eqref{lrk_h}. Therefore, the results for the long-range pairing Kitaev chain of Ref.~\onlinecite{Dutta2017} present a different structure of the excitation probability and they cannot be directly connected to the dynamics of LRTFIC.

The dynamical critical value $h_{c}^d$ of the final external field $h_{f}$ as a function of $\alpha$ is shown in Fig.~\ref{Fig2c}. The doubly critical phase shrinks with increasing $\alpha$, since it has to vanish in the $\alpha\to\infty$ limit. Then, the question arises whether this phase asymptotically vanishes in the latter limit or rather terminates abruptly at some value $\alpha_{\cross}$. Using tree-level scaling arguments, one can show\cite{Defenu2015} that long-range interactions cannot be relevant in LRTFIC for $\alpha\geq3$ while, in the case of LRKC, they turn out to be irrelevant also for $\alpha\geq2$, at least in equilibrium.\cite{Alecce2017,Lepori2016Gori} Therefore, one expects long-range interactions not to alter the universal behavior of the model beyond a certain threshold value $\alpha_{\cross}$ also in the dynamical realm. This is indeed the case, but the actual value of $\alpha_{\cross}$ extends well beyond the expected equilibrium regimes, as shown in the following.

%
%

\begin{figure}[h!]
	\includegraphics[width=.45\textwidth]{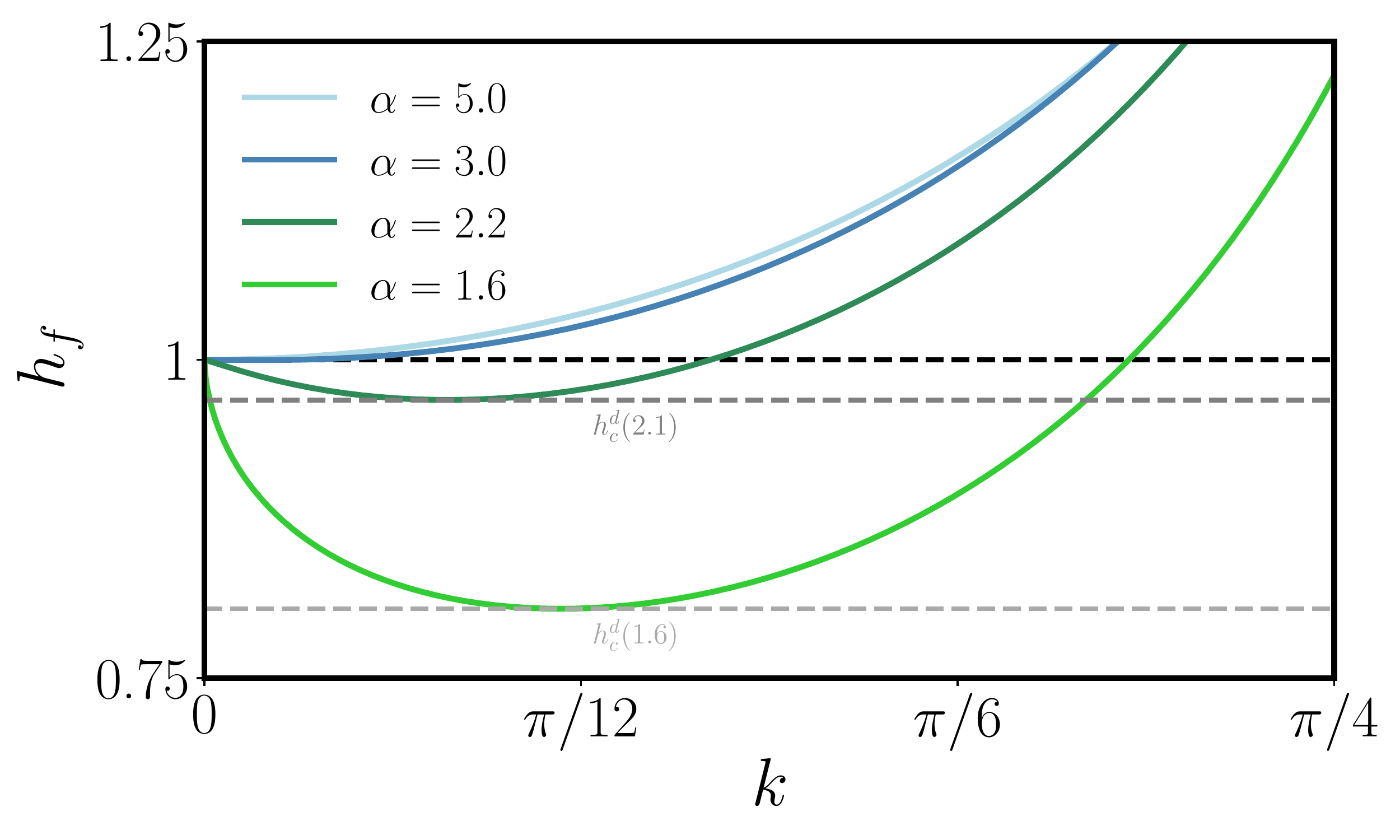}
	\caption{\label{Fig4} The final magnetic field at which one observes critical modes as a function of their momentum $k$ for different values of the range parameter $\alpha=1.6$, $2.2$, $3.0$, and $5.0$ from bottom to top, respectively, in the $h_{i}=0$ case. All the curves start at $h_{f}=h_{c}^{e}=1$ for $k=0$ signaling the emergence of the singly critical regular dynamical phase for quenches from $h_i=0$ to $h_{f}>h_{c}^{e}$. For $\alpha<3$ the curves show a negative (nonanalytic) slope signaling the emergence of dynamical critical modes even for $h_{f}<h_{c}^{e}$ at $k>0$. At larger $k$, the $\alpha<3$ curves increase again, leading to two dynamical critical momenta for each $h_{f}$. Therefore, the $\alpha<3$ curves support a doubly critical regular dynamical phase and $h_{c}^{d}$ can be obtained as the minimum $h_{f}$ which supports critical modes, as shown by the dashed gray lines. For $\alpha>3$ the curves may still present negative curvature, and then support the doubly critical phase depending on the value of $h_{i}$. In the $h_{i}=0$ case, shown in the plot, the doubly critical phase ceases to exist for $\alpha\geq\alpha_{\cross}\approx3.35$, where the curves acquire positive curvature, regardless of the value of $h_i$.}
\end{figure}

Let us now investigate in detail the behavior of $h_{c}^{d}$ as a function of $\alpha$ and in particular whether the doubly critical dynamical phase collapses with the equilibrium criticality at a finite value of $\alpha$. It is convenient to directly consider the more general case of quenches starting at an initial point $h_{i}<h_c^e=1$ and rewrite~\eqref{kstar_eq} as
\begin{align}
\label{hfk}
h^{*}_{f}(k)=j_{k}+\frac{\Delta_{k}^{2}}{j_{k}-h_{i}},
\end{align}
which defines the values of $h_{f}=h^{*}_{f}(k)$ at which one has a dynamical critical mode with momentum $k$. The resulting curves, in the $h_{i}=0$ case, are shown in Fig.~\ref{Fig4}.  Each line starts from the equilibrium critical value $h_{c}^{e}=1$ at $k=0$, showing that the dynamical critical mode for a quench at the equilibrium phase boundary is exactly the equilibrium critical mode. For large $\alpha$ the curves are monotonically increasing and the lowest possible critical final field remains the one associated with the equilibrium critical mode $k=0$, with $h^{*}_{f}(k=0)=1$. However, for smaller $\alpha$ the $h^{*}_{f}(k)$ curves initially decrease, allowing for the existence of a dynamical critical phase for $h_{f}<h_c^e$, while $h_{f}=h_c^e$ always remains the solution at $k=0$ since the equilibrium phase diagram does not depend on $\alpha$ in our system; see lower green curves in Fig.~\ref{Fig4}.  At larger $k$ values the $h^{*}_{f}(k)$ curves bend upwards, leading to two $k$ values for $h_{f}<h_c^e$ thus indicating the emergence of the doubly critical regular dynamical phase. Changing the initial field $h_{i}$ of the quench does not influence this qualitative picture. One may expect that the doubly critical dynamical phase disappears for $\alpha\geq 3$, since nonanalytic corrections in the hopping term $j_{k}$, see~\eqref{hfk}, become subleading in this limit. However, the situation is in fact more subtle.

In order to identify the upper boundary of the doubly critical dynamical phase, one shall consider the low-momentum limit of~\eqref{hfk} in the neighborhood of $\alpha=3$:
\begin{align}
\label{hfk_low}
\lim_{k\to0}h^{*}_{f}(k)=\begin{cases}&1+a\,k^{\alpha-1}+\mathscr{O}(k^{\alpha}),\,\,\,\,\mathrm{if}~\alpha<3,\\
&1+c\,k^{2}+\mathscr{O}(k^{\alpha-1}),\,\,\,\,\,\mathrm{if}~\alpha\geq3,\end{cases}
\end{align}
where subleading logarithmic corrections at $\alpha=3$ have been neglected. The explicit expressions for the coefficients are
\begin{align}
a&=\frac{\sin\left(\frac{\alpha\pi}{2}\right)\Gamma(1-\alpha)}{\zeta(\alpha)},\\
c&=\frac{1}{1-h_{i}}\frac{\zeta^2(\alpha-1)}{\zeta^2(\alpha)}-\frac{\zeta(\alpha-2)}{2\zeta(\alpha)}.
\end{align}
The fact that the low-energy limit of $h^{*}_{f}(k)$ changes at the $\alpha=3$ boundary, see~\eqref{hfk_low}, suggests possible modifications in the universal behavior of the model. Nevertheless, the existence of the doubly critical dynamical phase is related to the curvature of $h^{*}_{f}(k)$ and, therefore, to the sign of the $a$ and $c$ coefficients in the expansion~\eqref{hfk_low}. For $\alpha<3$ one has $a<0$ independently of the values of $h_{i}$, and thus the doubly critical dynamical phase always exists in this region independently of $h_{i}$ and of any modification of the irrelevant contributions to the coupling terms $\Delta_{k}$ and $j_{k}$. These findings are in agreement with what is expected for LRTFIC, where power-law interactions are relevant for $\alpha<3$, which surprisingly exceeds the equilibrium critical threshold $\alpha=2$ for LRKC. On the other hand, for $\alpha\geq 3$ the coefficient $c$ changes sign as a function of $\alpha$ and $h_{i}$, indicating that the doubly critical dynamical phase may or may not exist depending on the microscopic details of the model and initial state. The maximum extension of the doubly critical phase is obtained in the $h_{i}=0$ case, which gives $\alpha_{\cross}\approx3.35$.

The emerging landscape for the dynamical critical properties of LRKC with power-law decaying interactions is quite involved: for $\alpha<3$ the phase diagram always contains two regular phases, with one that is singly critical for quenches across the equilibrium critical point $h_{c}^{e}=1$ independently of $\alpha$, and with the second that is doubly critical, which is found for quenches across the $h_{c}^{d}$ boundary but before $h_c^e$. For $\alpha>3$, where power-law couplings are irrelevant in equilibrium in both LRKC and LRTFIC, the doubly critical dynamical phase may still be found depending on the initial field up to an upper critical value of $\alpha_{\cross}\approx3.35$. However, this feature of the $\alpha>3$ phase diagram is not universal and may be modified by the addition of other irrelevant contributions to the interaction term.
	\subsection{Exponential couplings}
\begin{figure*}[t!]
		\centering
		\subfigure[$\quad h_f<h_{c}^e$]{\label{Fig3a}\includegraphics[width=.32\textwidth]{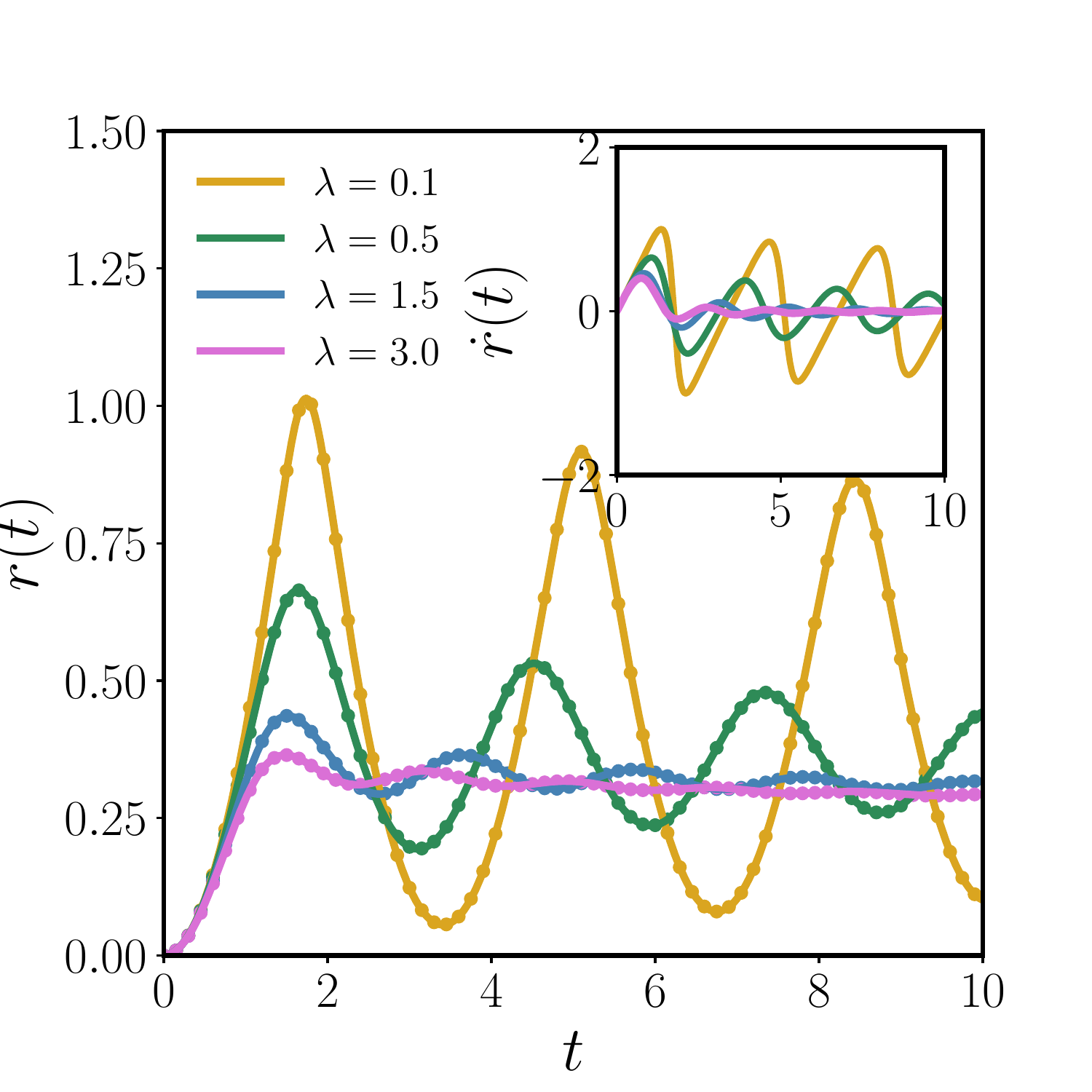}}
		\hfill
		\subfigure[$\quad h_f>h_{c}^e$]{\label{Fig3b}\includegraphics[width=.32\textwidth]{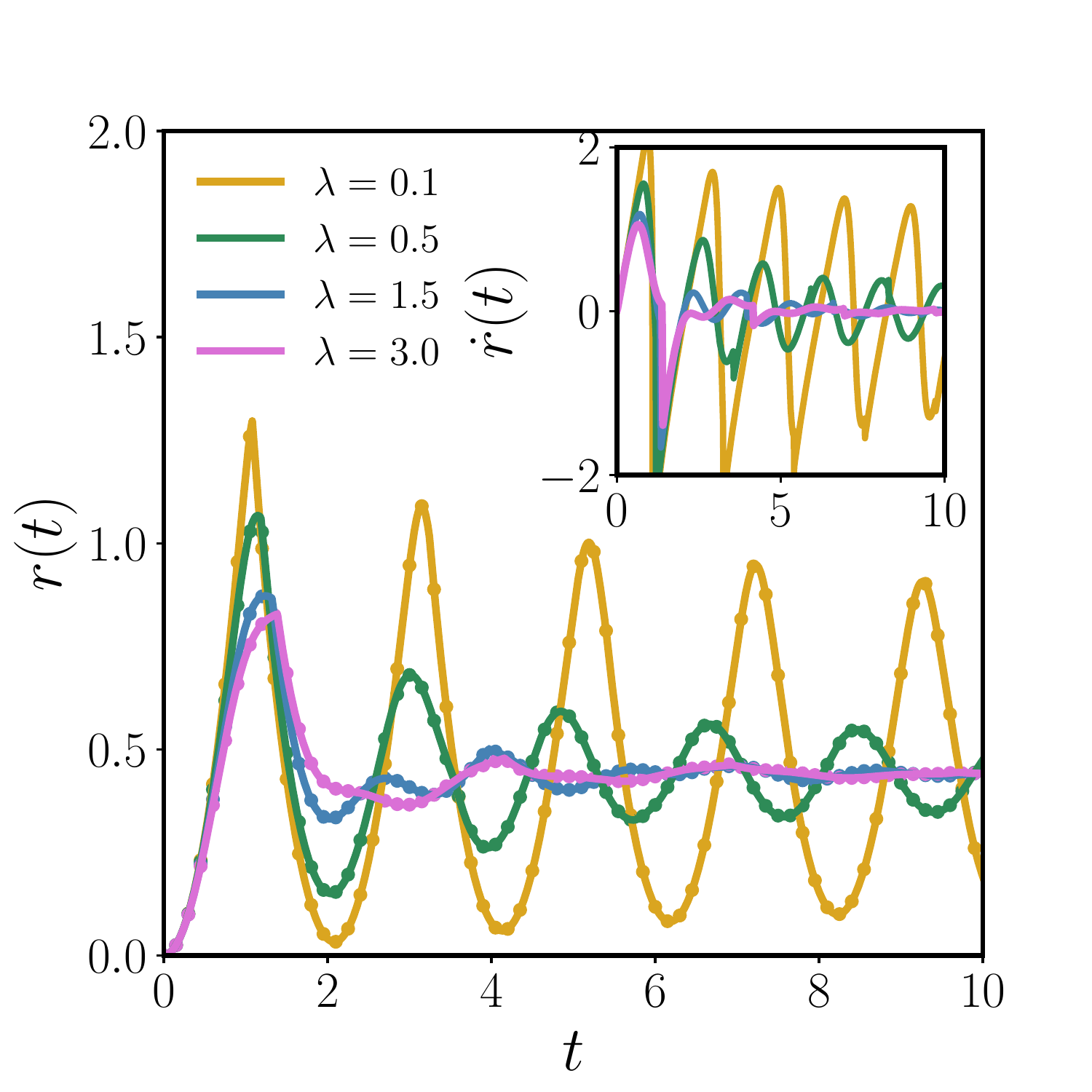}}
		\hfill
		\subfigure[$\quad p_{k}$]{\label{Fig3c}\includegraphics[width=.32\textwidth]{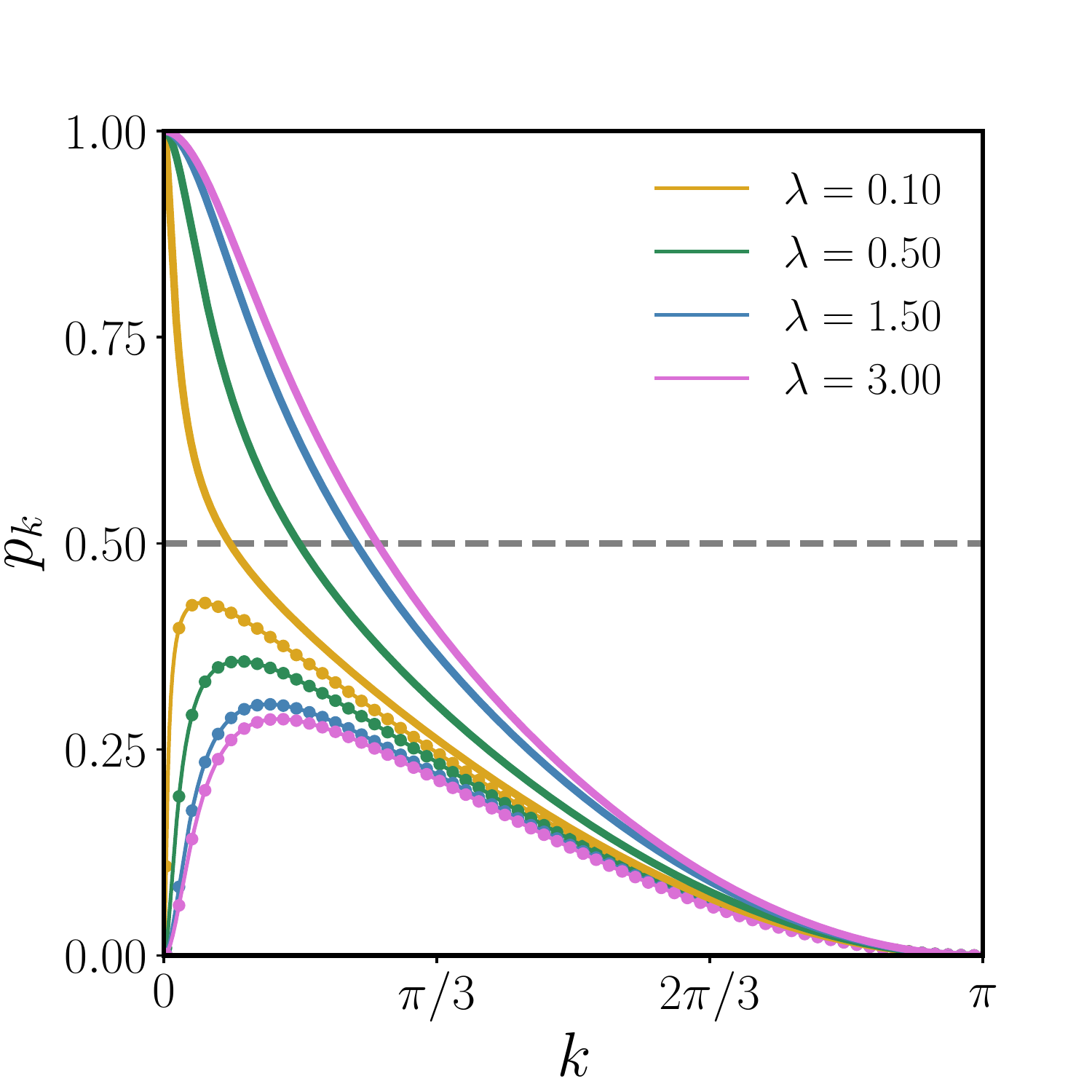}}
		\caption{\label{Fig3} 
			In panels (a) and (b) we show the return rates for quenches from $h_i=0$ to $h_{f}=0.9$ and $1.5$, respectively, for LRKC with exponentially decaying hopping and pairing. Cusps in the return rate only appear for $h_{f}>h_{c}^e$, where $h_{c}^e=1$ is the equilibrium (and dynamical) critical point for all $\lambda$. The solid lines represent the chain in the thermodynamic limit, while the dots are the finite-size system with periodic boundary conditions. In the case of exponentially decaying couplings no additional dynamical critical phase appears in LRKC, in contrast to the case of power-law couplings; cf.~Fig.~\ref{Fig2}. Indeed, the excitation probabilities in panel (c) do not cross the $1/2$ values for quenches to $h_{f}=0.9$ (solid lines with points) at any finite $k$. Only for quenches across the equilibrium phase transition $h_{f}>1$, which are indicated by solid lines in panel (c), does the excitation probability cross the $1/2$ threshold at a finite $k$ giving rise to cusps in the return rates shown in panel (b). Insets in panels (a) and (b) show the time derivative of the return rate, which shows jumps at the critical times for panel (b).}
	\end{figure*}

	The above picture is rather simplified if one considers exponentially decaying interactions $V_{|l-j|}=\exp[-\lambda(|l-j|-1)]/\mathcal{N}_\lambda$, with $\mathcal{N}_\lambda$ a normalization to ensure $h_c^e=1$ for all values of $\lambda$. Such an interaction profile places LRTFIC in the short-range equilibrium universality class for each $\lambda>0$, but nevertheless leads to a very rich dynamical phase diagram that exhibits criticality vastly distinct from that of NNTFIC.\cite{Halimeh2018a} Indeed, LRTFIC with exponentially decaying interactions always shows anomalous cusps in the return rate for sufficiently small values of $h_f$ at which $\hat{H}_f$ has bound domain walls in its spectrum. However, upon a truncated JW transformation into LRKC with exponentially decaying hopping and pairing, the anomalous cusps are absent, and the dynamical criticality is similar to that of NNTFIC with two dynamical phases where only quenches across the equilibrium critical point lead to cusps at a single critical momentum mode; cf.~Fig.~\ref{Fig3}. Indeed, even though in the case of LRTFIC there is a rich dynamical phase diagram hosting anomalous and regular phases in addition to a coexistence region\cite{Halimeh2018a} of both for $\lambda\lesssim0.6$, in the case of LRKC only the singly critical regular phase and the trivial phase exist. This shows how crucial the effect of domain-wall coupling is on DQPT as its presence leads to drastically different dynamical criticality that is completely absent in LRKC.

	Moreover, it is interesting to note that even though LRKC has no domain-wall coupling regardless of whether interactions are power-law or exponentially decaying, both types of connectivity lead to very different criticality, with power-law profiles leading to a much richer dynamical phase diagram.
	
	
	\section{Domain-wall picture}\label{sec:dw}
	The fermionic operators describe kink and anti-kink excitations of the initial spin Hamiltonian. These can be easily understood by considering their inverse JW transformation
	\begin{align}
		\hat{c}_{j}&=-\frac{1}{2}\hat{K}_j\big(\hat{\sigma}_j^z-i\hat{\sigma}^{y}_{j}\big),
		\end{align}
		with the kink operator defined as
		\begin{align}
	\hat{K}_{j}&=\prod_{m=1}^{j-1}\hat{\sigma}^x_m.
	\end{align} 
Therefore, the fermionic operators in spin language correspond to creating domain walls.\cite{Fradkin2013} To see this, let us choose the representation $\{\ket{\uparrow},\ket{\downarrow}\}$ for the eigenstates of $\hat{\sigma}^{z}$ and consider the ground state of the system at $h=0$, which reads
	\begin{align}
	\ket{0}_{h=0}=\ket{\uparrow}\otimes\cdots\otimes\ket{\uparrow},
	\end{align}
	where the $z$-up polarized state has been chosen for defineteness.  The kink operator acts as
	\begin{align}
	\label{s12}
	\hat{K}_{j}\ket{0}_{h=0}=\ket{\downarrow}\otimes\cdots\otimes\ket{\downarrow}_{j-1}\otimes\ket{\uparrow}_{j}\otimes\cdots\otimes\ket{\uparrow},
	\end{align} 
	and thus a \textit{sharp} domain wall has been created in the system. The sharp domain wall state has a discontinuity in the $m_{j}=\langle\hat{\sigma}^{z}_j\rangle$ magnetization, which jumps from $-1$ to $1$ from site $j-1$ to site $j$. In contrast, the fermionic operator $\hat{c}_{j}$ generates a \textit{smooth} domain wall,
	\begin{align}\nonumber
	&\hat{c}_j\ket{0}_{h=0}=-\frac{1}{2}\hat{K}_j\big(\hat{\sigma}_j^z-i\hat{\sigma}^{y}_{j}\big)\ket{0}_{h=0}\\\nonumber
	=&-\frac{1}{2}\ket{\downarrow}\otimes\cdots\otimes\ket{\downarrow}_{j-1}\otimes\ket{\uparrow}_{j}\otimes\ket{\uparrow}_{j+1}\otimes\cdots\otimes\ket{\uparrow}\\\nonumber
	&-\frac{1}{2}\ket{\downarrow}\otimes\cdots\otimes\ket{\downarrow}_{j-1}\otimes\ket{\downarrow}_{j}\otimes\ket{\uparrow}_{j+1}\otimes\cdots\otimes\ket{\uparrow}\\
	=&-\frac{1}{\sqrt{2}}\ket{\downarrow}\otimes\cdots\otimes\ket{\downarrow}_{j-1}\otimes\ket{+}_{j}\otimes\ket{\uparrow}_{j+1}\otimes\cdots\otimes\ket{\uparrow},
	\end{align} 
	localized exactly at site $j$, where we adopt the notation $\ket{+}=(\ket{\uparrow}+\ket{\downarrow})/\sqrt{2}$. Therefore, the $\hat{c}_{j}$ operators represent smooth domain wall excitations, which are perfectly localized in space and can travel around the system. A smooth domain wall can be seen as the coherent superposition of two sharp domain walls at neighboring positions. Moreover, since the Hamiltonian of LRTFIC commutes with the parity operator $\hat{P}=\prod_{i}^{N}\hat{\sigma}^{x}_{i}$, the overall parity of the system is conserved, and, therefore, only double-domain wall excitations (with different parities) are allowed upon time-dependent variation of each of the system parameters.

	It is now straightforward to interpret the physics of LRTFIC. For nearest-neighbor interactions the domain-wall excitations freely propagate in the system, since they do not interact with each other. In contrast, as soon as the couplings extend beyond nearest-neighbor sites, quartic terms appear in the fermionic representation of LRTFIC Hamiltonian; see~\eqref{eq:quartic}. Therefore, coherent domain wall states at different positions can collide with each other, leading to an interacting fermion system. This is not surprising since it is well known that as the connectivity increases for $\alpha\to 0$, the system tends towards its fully connected mean-field solution, which is exactly described in terms of bosonic spin-wave excitations~\cite{Lerose2018} and not in terms of free fermions.
	
	Upon performing the truncated JW transformation that leads to LRKC, the quartic fermionic terms in~\eqref{eq:quartic} are removed, thus eliminating domain-wall collisions. Therefore, we once again have a situation similar to that of NNTFIC where domain walls are freely propagating and the system maintains its purely fermionic character at all $\alpha$ values. The fact that this leads to the absence of anomalous dynamical criticality shows that domain-wall coupling is a necessary condition for anomalous cusps to appear in the return rate. Nevertheless, and as already mentioned, power-law couplings lead to an additional dynamical phase that hosts regular cusps at two critical momenta for $\alpha<3$, in addition to the dynamical phase also present in NNTFIC where only one critical momentum leads to cusps.
	
	\section{Conclusions}\label{sec:conc}
	In summary, we have shown that when coupling between domain walls is switched off, anomalous dynamical criticality vanishes, and DQPT behavior is similar to that of the nearest-neighbor transverse-field Ising chain. In particular, the long-range transverse-field Ising chain with either power-law or exponentially decaying interactions, which hosts anomalous dynamical criticality when the quench Hamiltonian has bound domain walls in its spectrum, is mapped onto a quadratic fermionic system using the truncated JW transformation. The \textit{truncation} lies in removing higher-than-quadratic fermionic terms in the fermionic model that one achieves upon a full JW transformation. Since we show that the quartic fermionic term describes coupling between domain walls in the spin picture, the absence of anomalous criticality upon the removal of this term indicates that domain-wall coupling is a necessary condition for anomalous cusps to appear in the return rate. This finding signifies the importance of the dynamics of domain walls in spin systems, particularly when it comes to emergent criticality in the wake of quench protocols.
	
	Focusing on the Kitaev chain with extensive long-range hopping and pairing terms, we find a rich dynamical phase diagram in the case of power-law profiles. For sufficiently small power-law exponent, a doubly critical regular dynamical phase appears in the return rate for final quench-parameter values between the dynamical and equilibrium critical points. This phase is characterized by two sets of cusps, each of which has its own critical momentum mode and temporal periodicity. For larger power-law exponents, and for all nonzero exponential-decay exponents, DQPT behavior is identical to that of the nearest-neighbor case where only two (rather than three) dynamical phases occur: a singly critical regular phase for quenches across the equilibrium critical point that shows cusps in the return rate originating from one critical momentum and occurring periodically in time, and a trivial phase for quenches within the same equilibrium phase where the return rate is fully analytic.
	
The existence of the doubly critical phase for LRKC with power-law couplings is shown to depend only on the universal low-energy properties of the model for $\alpha<3$.  This result is already surprising since this phase includes the region $2<\alpha<3$ where power-law couplings are irrelevant for the equilibrium phase diagram. This unexpected feature is the result of major sensitivity of dynamical criticality to long-range corrections in the hopping term with respect to the equilibrium behavior. Notably, this region where power-law interactions are relevant for dynamical criticality agrees with the one of LRTFIC, where power-law couplings remain relevant till $\alpha\approx3$ also in equilibrium. However, for $\alpha>3$ the doubly critical phase does not univocally disappear, but may or may not exist depending on the initial field $h_{i}$. Even more importantly, the existence of this phase for $\alpha>3$ is related to the curvature of $h_{f}^{*}(k)$, see~\eqref{hfk}, which, in turn, may be altered by the addition of irrelevant interaction terms in Hamiltonian~\eqref{eq:approx}, thereby leading to the failure of universality in this regime. This rich behavior suggests that the exploration of dynamical criticality in long-range interacting models can shed new light also on the equilibrium universal scaling close to interacting quantum critical points.
	
	\section*{Acknowledgments}
	The authors are grateful to Utso Bhattacharya, Amit Dutta, Fabian H.~L.~Essler, Giovanna Morigi, Maarten Van Damme, and Laurens Vanderstraeten for stimulating discussions. This work is part of the DFG Collaborative Research Centre SFB 1225 ISOQUANT, as well as the Cluster of Excellence STRUCTURES at the University of Heidelberg.
\appendix

	\section{Bogoliubov Transformation}
	\label{AppB}
	The Bogoliubov transformation, which diagonalizes LRKC in the thermodynamics limit, is described here in detail. Starting with the real-space Hamiltonian~\eqref{lrk_h} we analytically solve the model by first introducing a Fourier transformation for the fermionic operators $\hat{c}_l,\hat{c}_l^\dagger$. This is possible for periodic or antiperiodic boundary conditions (PBC or ABC). Unfortunately, the long-range couplings of~\eqref{lrk_h} prevent the direct application of PBC (ABC) since they would cancel the long-range hopping (pairing) terms due to fermionic anti-commutation rules. In order to analytically study the thermodynamic limit, it is convenient to introduce a regularized version of~\eqref{lrk_h}, where long-range couplings have been truncated as follows:\cite{Alecce2017}
	\begin{align}
	\hat{\mathcal{H}}_{N}=&\,-\sum_{l}\Bigg\{\sum_{r=1}^{\frac{N}{2}-1}V_r\big(\hat{c}_l^\dagger\hat{c}_{l+r}+\hat{c}_l^\dagger\hat{c}_{l+r}^\dagger-\hat{c}_l\hat{c}_{l+r}-\hat{c}_l\hat{c}_{l+r}^\dagger\big)\nonumber\\
	&-h\big(1-2\hat{c}_l^\dagger\hat{c}_l\big)
	\Bigg\},\label{lrk_h_reg}
	\end{align}
	where $V_r$ represents the real-space couplings as a function of the site distance $r$, as can be done for translation invariant $V_{|l-j|}$. In the thermodynamic limit $N\to\infty$ the two Hamiltonians~\eqref{lrk_h} and~\eqref{lrk_h_reg} coincide, and they shall exhibit the same critical behavior for $\alpha>1$ when boundary conditions are expected to be irrelevant for infinite sizes. The long-range couplings are normalised by
		\begin{align}
	\label{coupl_val}
	\mathcal{N}_{\alpha,\lambda}=\sum_{r=1}^{\frac{N}{2}-1}V_r,
	\end{align}
	where the subscript indicates whether the coupling is power law ($\alpha$) or exponential decay ($\lambda$).

	Hamiltonian\,\eqref{lrk_h_reg} supports PBC at each $N$ and we can safely consider the Fourier opearators,
	\begin{align}
	\label{f_trans}
	\hat{c}_{k}=e^{-i\frac{\pi}{4}}\sum_{r=1}^{N}\hat{c}_{r}e^{ikr},
	\end{align}
	where $k\equiv2\pi n/N$ and $n\in\{0,1,\ldots,N-1\}$ since we choose PBC, even though it is to be noted that using ABC changes nothing in the thermodynamic limit. Plugging~\eqref{f_trans} into~\eqref{lrk_h_reg}, one finds
	\begin{align}
	\label{h_klr_app}
	H=\sum_{k}\Big[(\hat{c}^{\dagger}_{k}\hat{c}_{k}
	-\hat{c}_{-k}\hat{c}^{\dagger}_{-k})\varepsilon_{k}+(\hat{c}^{\dagger}_{k}\hat{c}^{\dagger}_{-k}+\hat{c}_{-k}\hat{c}_{k})\Delta_{k}\Big],
	\end{align}
	apart from an inconsequential constant energy term that has been neglected. The Fourier-space couplings are listed below, but before specifying them it is convenient to describe the general diagonalization procedure of~\eqref{h_klr}. First of all, one has to employ a Bogoliubov transformation
	\begin{align}\label{eq:Bogo}
	\hat{c}_{k}=u_{k}\hat{\gamma}_{k}+v^{*}_{-k}\hat{\gamma}^{\dagger}_{-k}
	\end{align}
	in order to diagonalize the static Hamiltonian, where the Bogoliubov operators $\hat{\gamma}_{k}$ and $\hat{\gamma}_{k}^\dagger$ satisfy the fermionic anticommutation relations. The transformation~\eqref{eq:Bogo} casts the Hamiltonian in diagonal form
	\begin{align}
	\hat{H}=\sum_{k}\,\omega_{k}\left(\hat{\gamma}^{\dagger}_{k}\hat{\gamma}_{k}-\frac{1}{2}\right),
	\end{align}
	provided that 
	\begin{align}
	(u_{k},v_{k})=\left(\cos\frac{\theta_{k}}{2}, \sin\frac{\theta_{k}}{2}\right),
	\end{align}
	with
	\begin{align}
	\tan\theta_{k}=\frac{\Delta_k}{\varepsilon_k},
	\end{align}
	and  eigenfrequencies
	\begin{align}
	\label{spec_eq}
	\omega_{k}=\sqrt{\varepsilon_k^{2}+\Delta_k^{2}}.
	\end{align}
	This transformation solves the equilibrium model.
	We are now ready to explicitly consider the Fourier-space couplings for our system. The Fourier couplings in~\eqref{h_klr_app} read
	\begin{align}
	\varepsilon_k&=h-j_k,\label{e_tf}\\
	j_k&=\frac{1}{\mathcal{N}_{\alpha,\lambda}}\sum_{r=1}^{\frac{N}{2}-1}V_r\cos(kr),\label{kin_tf_N}\\
	\Delta_k&=\frac{1}{\mathcal{N}_{\alpha,\lambda}}\sum_{r=1}^{\frac{N}{2}-1}V_r\sin(kr).\label{par_tf_N}
	\end{align}
	
	\begin{figure}[h!]
		\includegraphics[width=.45\textwidth]{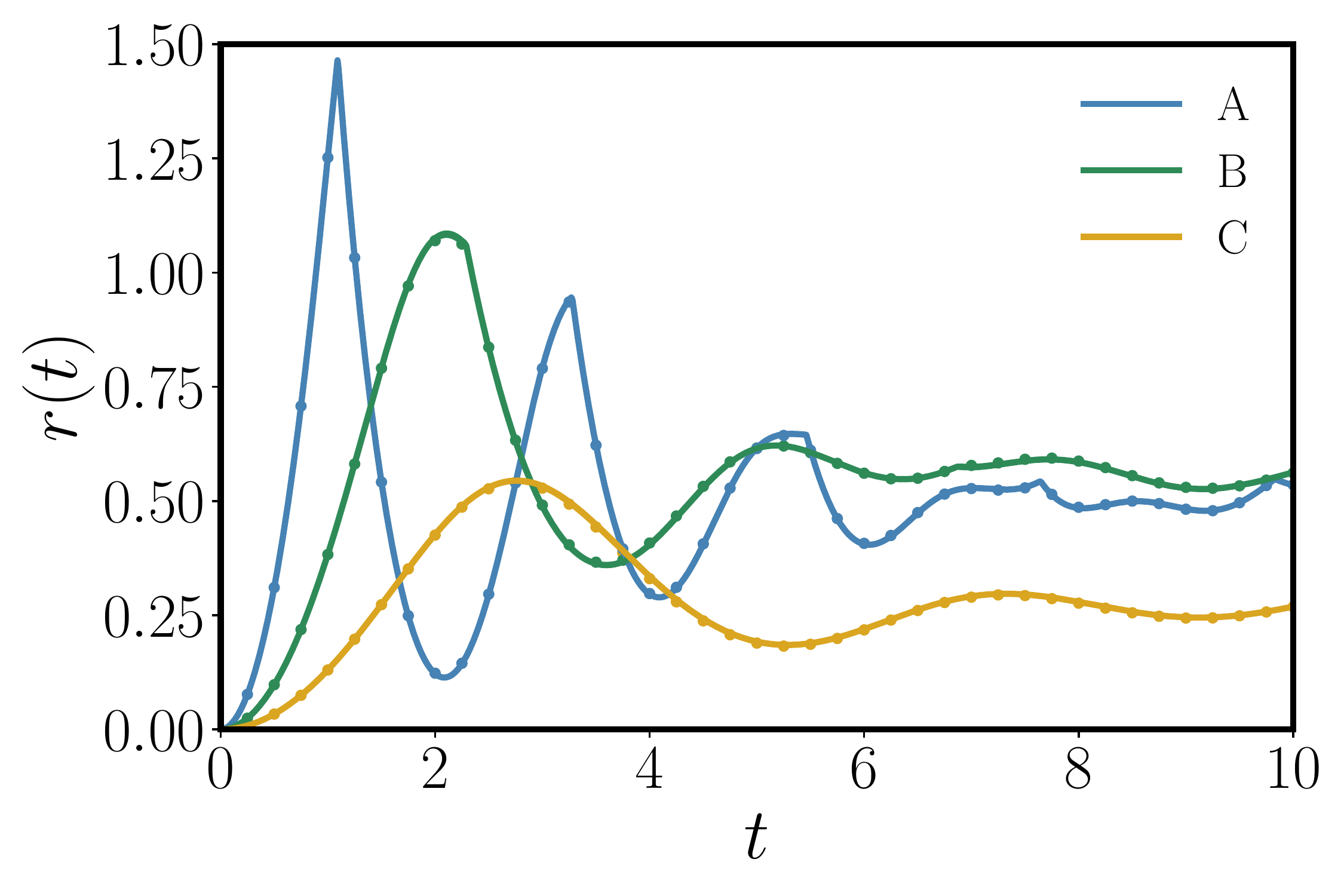}
		\caption{\label{Fig1app} Return rate $r(t)$ at $\alpha=1.5$ for $N=50000$ compared to the solution obtained by~\eqref{exp_ret_rate} using the analytical expressions~\eqref{kin_tf} and~\eqref{par_tf} in the thermodynamic limit.} 
	\end{figure} 
	\subsection{Power-law couplings}
	For $\alpha>1$ the normalization only introduces a finite coefficient $N_{\alpha}=\zeta(\alpha)$ in the thermodynamic limit, which fixes the equilibrium critical point of the model at $h_{c}^e=\pm1$ irrespective of the value of $\alpha$. 
	In the thermodynamic limit one can directly consider the $N\to\infty$ limit of the above expressions: 
	
	\begin{align}
	j_k&=\frac{1}{\zeta(\alpha)}\sum_{r=1}^{\infty}\frac{\cos(kr)}{r^{\alpha}}=\frac{\mathrm{Re[\,Li}\left(e^{ik}\right)]}{2\zeta(\alpha)},\label{kin_tf}\\
	\Delta_k&=\frac{1}{\zeta(\alpha)}\sum_{r=1}^{\infty}\frac{\sin(kr)}{r^{\alpha}}=\frac{\mathrm{Im[\,Li}\left(e^{ik}\right)]}{2\zeta(\alpha)},\label{par_tf}
	\end{align}
	where the $\zeta(\alpha)$ normalization forces the constant coefficient in the Fourier expansion of~\eqref{kin_tf} to be unity and the momentum now takes  continuous values $k\in[-\pi,\pi]$. These expressions enable the exact solution of~\eqref{h_klr_app}, which represents the thermodynamic limit of both Hamiltonians in Eqs.~\eqref{lrk_h} and~\eqref{lrk_h_reg}. Indeed, numerical solutions of the finite chain in~\eqref{lrk_h_reg} for large $N$ and of the thermodynamic-limit Hamiltonian are in very good agreement; see Fig.~\ref{Fig1}. It is worth noting that the thermodynamic limit becomes more difficult the closer $\alpha$ is to the divergent limit $\alpha\to1$. Indeed for $N=1000$, the dotted curves collapse onto the thermodynamic-limit solution (solid curves) in Fig.~\ref{Fig1b} and~\ref{Fig1c}, for $\alpha=2.2$ and $5.5$ respectively. In contrast, for $\alpha=1.5$ the finite-chain solutions (dotted curves) at $N=1000$ do not perfectly agree with the thermodynamic-limit solutions (solid curves) in Fig.~\ref{Fig1a}, but very good agreement can be nevertheless attained for larger $N$ as shown in Fig.~\ref{Fig1app}, where the periodic-chain solution has been carried out for $N=50000$.
	
	\subsection{Exponential couplings}
	In the case of exponential couplings $V_{|l-j|}=e^{-\lambda(|l-j|-1)}/\mathcal{N}_\lambda$, the Fourier-space couplings in the thermodynamic limit are easily written as
	\begin{align}
	j_k&=-\frac{e^{\lambda}}{2\mathcal{N}_{\lambda}}\left(1-\frac{\sinh\lambda}{\cosh\lambda-\cos k}\right),\label{kin_tf_exp}\\
	\Delta_k&=\frac{e^{\lambda}}{2\mathcal{N}_{\lambda}}\frac{\sin k}{\cosh\lambda-\cos k},\label{par_tf_exp}
	\end{align}
	where $\mathcal{N}_{\lambda}=\frac{e^{\lambda}}{e^{\lambda}-1}$.
	\bibliography{LRKC_biblio}

\begin{thebibliography}{58}
\bibitem[{foo({\natexlab{b}})}]{footnote3}
\bibinfo{howpublished}{By `long-range' we do not necessarily mean in the
  renormalization-group sense of leading to a finite-temperature phase
  transition, but rather that the interactions are spread over an extensive
  part of the chain.}
\end{thebibliography}
\end{document}